\documentclass{ws-p8-50x6-00}

\def\p{\partial}
\def\n{\nabla}
\def\d{\displaystyle}

\begin{document}

\title{Gravitational Interaction of Higher Spin Massive Fields
and String Theory}

\author{I.L. Buchbinder}

\address{
Instituto de F\'\i sica, Universidade de S\~ao Paulo,\\
P.O. Box 66318, 05315-970, S\~ao Paulo, SP, Brasil\\
Department of Theoretical Physics,
Tomsk State Pedagogical University,\\
Tomsk 634041, Russia
\\E-mail: joseph@tspu.edu.ru}

\author{V.D. Pershin}

\address{
Department of Theoretical Physics,
Tomsk State University,\\
Tomsk 634050, Russia \\
E-mail: pershin@ic.tsu.ru}

\maketitle

\abstracts{
We discuss the problem of consistent description of higher spin
massive fields coupled to external gravity. As an example we consider
massive field of spin 2 in arbitrary gravitational field. Consistency
requires the theory to have the same number of degrees of freedom as
in flat spacetime and to describe causal propagation. By careful
analysis of lagrangian structure of the theory and its constraints we
show that there exist at least two possibilities of achieving
consistency. The first possibility is provided by a lagrangian
on specific manifolds such as static or Einstein spacetimes.
The second possibility is realized in arbitrary curved
spacetime by a lagrangian representing an infinite series in
curvature. In the framework of string theory we derive equations of
motion for background massive spin 2 field coupled to gravity from
the requirement of quantum Weyl invariance. These equations appear to
be a particular case of the general consistent equations obtained
from the field theory point of view.}

\section{Introduction}

Despite many years of intensive studies the construction of
consistent interacting theories of higher spin fields is still far
from completion. Consistency problems arise both in higher spin
field theories with self-interaction and in models of a single higher
spin field in non-trivial external background. In this contribution
we review a recent progress\cite{ours,causality} achieved in building
of a consistent theory of massive spin 2 field in external gravity
and in understanding about the way the string theory predicts
consistent equations of motion for such a system.

In general, there are two ways the interaction can spoil the
consistency of a higher spin fields theory.
Firstly, interaction may change the number of dynamical
degrees of freedom. For example, a massive field with spin $s$ in
$D=4$ Minkowski spacetime is described by a rank $s$ symmetric
traceless transverse tensor $\phi_{(\mu_1\ldots\mu_s)}$ satisfying
the mass shell condition:
\be
(\p^2-m^2) \phi_{\mu_1\ldots\mu_s} =0 {,} \qquad
\p^\mu \phi_{\mu\mu_1\ldots\mu_{s-1}} =0 {,} \qquad
\phi^\mu{}_{\mu\mu_1\ldots\mu_{s-2}} =0 {.}
\label{irrep}
\ee
To reproduce all these equations from a single lagrangian one needs
to introduce auxiliary fields $\chi_{\mu_1\ldots\mu_{s-2}}$,
$\chi_{\mu_1\ldots\mu_{s-3}}$, \ldots, $\chi$\cite{fierz,singh}.
These symmetric traceless fields vanish on shell but their presence
in the theory provides lagrangian description of the conditions
(\ref{irrep}).  In higher dimensional spacetimes there appear fields
of more complex tensor structure but general situation remains the
same, i.e.  lagrangian description always requires presence of
unphysical auxiliary degrees of freedom.

Namely these auxiliary fields create problems when one tries to turn
on interaction in the theory. Arbitrary interaction makes the
auxiliary fields dynamical thus increasing the number of degrees of
freedom. Usually these extra degrees of freedom are ghostlike and
should be considered as pathological. Requirement of absence of these
extra dynamical degrees of freedom imposes severe restrictions on the
possible interaction\cite{aragone,higuchi,bengt,ovrut,klish2}.

The other problem that may arise in higher spin fields theories is
connected with possible violation of causal properties. This problem
was first noted in the theory of spin 3/2 field in external
fields\cite{zwanziger} (see also the review\cite{zwanziger2} and a
recent discussion in\cite{3/2})

In general,
when one has a system of differential equations for a set of fields
$\phi^B$ (to be specific, let us say about second order equations)
\be
M_{AB}{}^{\mu\nu}\p_\mu \p_\nu \phi^B + \ldots = 0 {,} \qquad
\mu,\nu=0,\ldots,D-1
\ee
the following definitions are used. A characteristic matrix is the
matrix function of $D$ arguments $n_\mu$ built out of the
coefficients at the second derivatives in the equations:
$
M_{AB}(n) = M_{AB}{}^{\mu\nu} n_\mu n_\nu {.}
$
A characteristic equation is
$
\det M_{AB} (n) = 0 {.}
$
A characteristic surface is the surface $S(x)=const$ where
$\p_\mu S(x)=n_\mu$.
If for any $n_i$ ($i=1,\ldots,D-1$) all solutions of the
characteristic equation $n_0(n_i)$ are real then the system of
differential equations is called hyperbolic and describes
propagation of some wave processes. The hyperbolic system is called
causal if there is no timelike vectors among solutions $n_\mu$ of the
characteristic equations. Such a system describes propagation with a
velocity not exceeding the speed of light. If there exist timelike
solutions for $n_\mu$ then the corresponding characteristic surfaces
are spacelike and violate causality.

Turning on interaction in theories of higher spin fields in general
changes the characteristic matrix and there appears possibility of
superluminal propagation. Such a situation also should be considered
as pathological. Note that the requirement of causal behaviour is an
independent condition. Interaction with external fields may violate
causality even in a covariant theory with the correct number of
degrees of freedom\footnote{It is intersting that
dimensional reduction can provide in principle a very specific
non-minimal coupling to external fields which preserve
consistency and causality, see discussion for spin 3/2 field
in\cite{sivakumar}.}.

As an example where both these problem arise we consider the theory
of massive spin 2 field in external gravitational field in arbitrary
spacetime dimension. In Section~2 we describe the structure of
its lagrangian equations of motion and constraints
and demonstrate how correct number of degrees of freedom can be
achieved in a number of specific spacetimes. Namely, we consider two
examples - an arbitrary static spacetime and an Einstein spacetime.
In both cases there exists correct flat spacetime limit though in
static case there may be regions where propagation of some of the
spin 2 field comonents is acausal.  Another possibility of achieving
consistency is described in Section~3 where lagrangian equations of
motion for massive spin 2 field are constructed in form of infinite
series in curvature. These kinds of infinite series arise naturally
in string theory which contains an infinite tower of massive higher
spin excitations and so should also provide a consistent scheme for
description of higher spin fields interaction.  Section~4 is devoted
to open string theory in background of massless graviton and massive
spin 2 field.  As is well known\cite{cfmp} the requirement of quantum
Weyl invariance of two-dimensional $\sigma-$model coupled to massless
background fileds gives rise to effective equations of motion for
these fields.  In case of massive background fields the coresponding
$\sigma-$model action is non-renormalizable and should contain an
infinite number of terms but as was shown in\cite{ourold} a specific
structure of renormalization makes it possible to calculate all
$\beta-$functions pertaining in each perturbative order only finite
number of counterterms. In linear order the effective equations of
motion obtained that way were shown to be in agreement with canonical
analysis of the corresponding $\sigma-$model action\cite{toder}.  In
this contribution we show that string theory also gives consistent
equations of motion for massive spin 2 field interacting with gravity
which represent a particular case of general equations described in
the previous sections.

\section{Massive spin 2 field on specific manifolds}

Let us start with reminding the lagrangian structure of a free
massive spin 2 field. To find the complete set of constraints we
use the general lagrangian scheme\cite{gitman} which
is equivalent to the Dirac-Bergmann procedure in hamiltonian
formalism but for our purposes is simpler.
In the case of second class constraints (which is
relevant for massive higher spin fields) it consists in the following
steps.  If in a theory of some set of fields $\phi^A(x)$,
$A=1,\ldots,N$ the original lagrangian equations of motion define
only $r<N$ of the second time derivatives (``accelerations'')
$\ddot\phi^A$ then one can build $N-r$ primary constraints, i.e.
linear combinations of the equations of motion that do not contain
accelerations. Requirements of conservation in time of the primary
constraints either define some of the missing accelerations or lead
to new (secondary) constraints.  Then one demands conservation of the
secondary constraints and so on, until all the accelerations are
defined and the procedure closes up.

In the flat spacetime the massive spin 2 field is described
by symmetric transversal and traceless
tensor of the second rank $H_{\mu\nu}$ satisfying mass-shell
condition:
\begin{equation}
\Bigr(\partial^2-m^2\Bigl) H_{\mu\nu}=0 {,}\qquad
\partial^\mu H_{\mu\nu}=0 {,}\qquad
H^\mu{}_\mu=0 {.}
\label{irred}
\end{equation}
In higher dimensional spacetimes Poincare algebras have more than two
Casimir operators and so there are several different spins for $D>4$.
Talking about spin 2 massive field in arbitrary dimension we will
mean, as usual, that this field by definition satisfies the same
equations (\ref{irred}) as in $D=4$. After dimensional reduction to
$D=4$ such a field will describe massive spin two representation of
$D=4$ Poincare algebra plus infinite tower of Kaluza-Klein
descendants.

All the equations (\ref{irred}) can be derived from the
Fierz-Pauli action\cite{fierz}:
\begin{eqnarray}
S&=&\int\! d^D x \biggl\{ \frac{1}{4} \partial_\mu H \partial^\mu H
-\frac{1}{4} \partial_\mu H_{\nu\rho} \partial^\mu H^{\nu\rho}
-\frac{1}{2} \partial^\mu H_{\mu\nu} \partial^\nu H
+\frac{1}{2} \partial_\mu H_{\nu\rho} \partial^\rho H^{\nu\mu}
\nonumber\\&&
\qquad\qquad
{} - \frac{m^2}{4} H_{\mu\nu} H^{\mu\nu} + \frac{m^2}{4}  H^2
 \biggr\}
\label{actfield}
\end{eqnarray}
where $H=\eta^{\mu\nu} H_{\mu\nu}$.

Here the role of auxiliary field is played by the trace
$H=\eta^{\mu\nu} H_{\mu\nu}$. The equations of motion
\begin{eqnarray}
E_{\mu\nu}&=&\partial^2 H_{\mu\nu} - \eta_{\mu\nu} \partial^2 H +
\partial_\mu \partial_\nu H
+ \eta_{\mu\nu} \partial^\alpha \partial^\beta H_{\alpha\beta}
- \partial_\sigma \partial_\mu H^\sigma{}_\nu
- \partial_\sigma \partial_\nu H^\sigma{}_\mu
\nonumber
\\&&\qquad\qquad
{}-m^2 H_{\mu\nu} + m^2 H \eta_{\mu\nu} = 0
\label{flateq}
\end{eqnarray}
contain $D$ primary constraints (expressions without second time
derivatives $\ddot H_{\mu\nu}$):
\bea
E_{00} &=& \Delta H_{ii} - \p_i\p_j H_{ij} - m^2 H_{ii} \equiv
\varphi_0^{(1)} \approx 0
\\
E_{0i} &=& \Delta H_{0i} +\p_i\dot H_{kk} -\p_k\dot H_{ki} - \p_i\p_k
H_{0k} - m^2 H_{0i} \equiv \varphi_i^{(1)} \approx 0 {.}
\eea
The remaining equations of motion $E_{ij}=0$ allow to define the
accelerations $\ddot H_{ij}$ in terms of $\dot H_{\mu\nu}$ and
$H_{\mu\nu}$. The accelerations $\ddot H_{00}$, $\ddot H_{0i}$ cannot
be expressed from the equations directly.

Conditions of conservation of the primary constraints in time
$\dot E_{0\mu}\approx 0$ lead to $D$ secondary constraints. On-shell
they are equivalent to
\be
\varphi_\nu^{(2)} = \partial^\mu E_{\mu\nu} =
m^2 \partial_\nu H - m^2 \partial^\mu H_{\mu\nu} \approx 0
\label{con1}
\ee

Conservation of $\varphi_i^{(2)}$ defines $D-1$ accelerations
$\ddot H_{0i}$ and conservation of $\varphi_0^{(2)}$ gives another
one constraint. It is convenient to choose it in the covariant form
by adding suitable terms proportional to the equations of motion:
\begin{eqnarray}
\varphi^{(3)}=
\partial^\mu \partial^\nu E_{\mu\nu}
+ \frac{m^2}{D-2} \eta^{\mu\nu} E_{\mu\nu}
=  H  m^4 \frac{D-1}{D-2} \approx 0
\label{con2}
\end{eqnarray}
Conservation of $\varphi^{(3)}$ gives one more constraint on initial
values
\be
\varphi^{(4)} = - \dot H_{00} + \dot H_{kk} = \dot H \approx 0
\ee
and from the conservation of this last constraint the acceleration
$\ddot H_{00}$ is defined. Altogether there are $2D+2$ constraints on
the initial values of $\dot H_{\mu\nu}$ and $H_{\mu\nu}$.

Obviously, the equations of motion (\ref{irred}) are causal because
the characteristic equation
\be
\det M(n) = (n^2)^{D(D+1)/2}
\ee
has 2 multiply degenerate roots
\be
-n_0^2+n_i^2 = 0, \qquad n_0 = \pm \sqrt{n_i^2} {.}
\ee
which correspond to real null solutions for $n_\mu$. Note that
analysis of causality is possible only after calculation of all the
constraints. Original lagrangian equations of motion (\ref{flateq})
have degenerate characteristic matrix $\det M(n)\equiv 0$ and do not
allow to define propagation cones of the field $H_{\mu\nu}$.

Now if we want to construct a theory of massive spin 2 field on a
curved manifold we should provide the same number of
propagating degrees of freedom as in the flat case. It means
that new equations of motion $E_{\mu\nu}$ should lead to exactly
$2D+2$ constraints  and in the flat spacetime limit these constraints
should reduce to their flat counterparts. The important point here is
that consistency does not require any specific transformation
properties of constraints in curved spacetime. For example,
in flat case the constraints $\varphi^{(2)}_\mu$ form a Lorentz
vector but there is no reason to require their curved counterpart to
be a vector with respect to local Lorentz transformation. The only
conditions one should care of is that the total number of constraints
should conserve and that they should always be of the second class.
In massless higher spin fields theory  one should also require
conservation of the corresponding gauge algebra and achieving
consistency in that case is a more difficult task\cite{vasiliev}.

Generalizing (\ref{actfield}) to curved spacetime we should substitute
all derivatives by the covariant ones and also we can add
non-minimal terms containing curvature tensor with some
dimensionless coefficients in front of them.  As a result, the most
general action for massive spin 2 field in curved spacetime quadratic
in derivatives and consistent with the flat limit should have the
form\cite{aragone}:
\begin{eqnarray}&&
S=\int d^D x\sqrt{-G} \biggl\{ \frac{1}{4} \nabla_\mu H \nabla^\mu H
-\frac{1}{4} \nabla_\mu H_{\nu\rho} \nabla^\mu H^{\nu\rho}
-\frac{1}{2} \nabla^\mu H_{\mu\nu} \nabla^\nu H
\nonumber
\\&&
\qquad
{}+\frac{1}{2} \nabla_\mu H_{\nu\rho} \nabla^\rho H^{\nu\mu}
+\frac{a_1}{2} R H_{\alpha\beta} H^{\alpha\beta}
+\frac{a_2}{2} R H^2
+\frac{a_3}{2} R^{\mu\alpha\nu\beta} H_{\mu\nu} H_{\alpha\beta}
\nonumber
\\&&
\qquad
{}+\frac{a_4}{2} R^{\alpha\beta} H_{\alpha\sigma} H_\beta{}^\sigma
+\frac{a_5}{2} R^{\alpha\beta} H_{\alpha\beta} H
{}- \frac{m^2}{4} H_{\mu\nu} H^{\mu\nu} + \frac{m^2}{4} H^2
 \biggr\}
\label{genact}
\end{eqnarray}
where $a_1, \ldots a_5$ are so far arbitrary dimensionless
coefficients, $R^\mu{}_{\nu\lambda\kappa}=\partial_\lambda
\Gamma^\mu_{\nu\kappa} -\ldots$,
$R_{\mu\nu}=R^\lambda{}_{\mu\lambda\nu}$.

Equations of motion contain second time derivatives of $H_{\mu\nu}$
in the following way:
\bea
E_{00} &=& (G^{mn}-G_{00}G^{00}G^{mn}+G_{00}G^{0m}G^{0n}) \nabla_0
\nabla_0 H_{mn} + O(\nabla_0) {,}
\nonumber\\
E_{0i} &=&
(-G_{0i}G^{00}G^{mn} + G_{0i}G^{0m}G^{0n} - G^{0m}\delta^n_i)
\nabla_0 \nabla_0 H_{mn} + O(\nabla_0) {,}
\nonumber\\
E_{ij} &=&
(G^{00}\delta^m_i\delta^n_j - G_{ij}G^{00}G^{mn} +
G_{ij}G^{0m}G^{0n}) \nabla_0 \nabla_0 H_{mn} + O(\nabla_0)
\eea
So we see that accelerations $\ddot H_{00}$ and $\ddot H_{0i}$ again
(as in the flat case) do not enter the equations of motion while
accelerations $\ddot H_{ij}$ can be expressed through $\dot
H_{\mu\nu}$, $H_{\mu\nu}$ and their spatial derivatives.

There are $D$ linear combinations of the equations of motion which
do not contain second time derivatives and so represent primary
constraints of the theory:
\be
\varphi^{(1)}_\mu = E^0{}_\mu = G^{00} E_{0\mu} + G^{0j} E_{j\mu}
\label{phi1}
\ee
At the next step one should calculate time derivatives of these
constraints and define secondary ones. In order to do this in a
covariant form we add to the time derivative of
$\varphi^{(1)}_\mu$ a linear combination of equations of motion and
primary constraints and define the secondary constraints as follows:
\be
\varphi^{(2)}_\mu = \nabla^\alpha E_{\alpha\mu}
\label{phi2}
\ee
Conservation of these $D$ secondary constraints
should lead to one new constraint and to expressions for $D-1$
accelerations $\ddot H_{0i}$. This means that the constraints
(\ref{phi2}) should contain the first time derivatives $\dot
H_{0\mu}$ through the matrix $\hat \Phi_\mu{}^\nu$ built out of the
blocks $A$, $B^j$, $C_i$, $D_i{}^j$
\bea
\varphi^{(2)}_0 &=&
A \; \dot H_{00} + B^j \dot H_{0j} + \ldots
\nonumber\\
\varphi^{(2)}_i &=&
C_i \dot H_{00} + D_i{}^j \dot H_{0j} + \ldots
\label{velocities}
\eea
whose rank is equal to $D-1$.

In the flat spacetime we had the matrix block elements
\be
A = B^j = C_i = 0, \qquad D_i{}^j = m^2 \delta_i^j
\ee
while in the curved case the explicit form of these elements in
the constraints (\ref{phi2}) is:
\bea
A &=& R G^{00} (2a_1+2a_2) + R^{00} (a_4+a_5)
           + R^0{}_0 G^{00} (a_4+a_5-1)
\nonumber\\
B^j &=& m^2 G^{0j} + R G^{0j} (2a_1+4a_2) + 2a_3 R^{0j}{}_0{}^0
      + R^j{}_0 G^{00} (a_4-2)
\nonumber\\&&{}
      + R^{0j} (a_4+2a_5) + R^0{}_0 G^{0j} (a_4+2a_5)
\nonumber\\
C_i &=& R^0{}_i G^{00} (a_4+a_5-1)
\nonumber\\
D_i{}^j &=&{} - m^2 G^{00} \delta_i^j + 2a_1 RG^{00} \delta_i^j
 + 2a_3 R^{0j}{}_i{}^0 + a_4 R^{00} \delta_i^j
\nonumber\\&&{}
 + (a_4-2) R^j{}_i G^{00} + (a_4+2a_5) R^0_i G^{0j}
\label{elements}
\eea

At this stage the restrictions that consistency imposes on the
type of interaction reduce to the requirements that the above matrix
elements give
\be
\det\hat\Phi=0 {,} \qquad \det D_i{}^j \neq 0
\label{main}
\ee
When the gravitational background is arbitrary it is not
clear how to fulfill this condition by choosing some
specific values of non-minimal couplings $a_1$, \ldots $a_5$. For
example, requirement of vanishing of the elements $A$
and $C_i$ (\ref{elements}) would lead to contradictory equations
$a_4+a_5=0$, $a_4+a_5-1=0$.

But the consistency conditions (\ref{main}) can be fulfilled in a
number of specific gravitational background. Namely, any spacetime
which in some coordinates has
\be
R^0{}_i=0
\label{example}
\ee
provides such an example.
In such a spacetime $R^{00}=R^0{}_0G^{00}$ and choosing coefficients
$a_1+a_2=0$, $2a_4+2a_5=1$ we have the first column of the matrix
$\hat\Phi$ vanishing and so the conditions (\ref{main}) fulfilled.

As a first example where (\ref{example}) holds let us consider an
arbitrary static spacetime, i.e. a spacetime having a timelike
Killing vector and invariant with respect to the time reversal
$x^0\to -x^0$. In such a spacetime one can always find
coordinates where
\be
\p_0 G_{\mu\nu}=0 {,} \qquad  G_{0i}=0 {.}
\ee
The matrix elements (\ref{elements}) in this case become
\bea
 &&A=RG^{00} (2a_1+2a_2) + R^{00} (2a_4+2a_5-1) {,}
\nonumber\\{}
&& B^j = 0 {,} \qquad\qquad C_i = 0 {,}
\\
&& D_i{}^j = ({} -m^2 G^{00} + 2a_1 RG^{00} + a_4 R^{00}) \delta_i^j
 + (a_4-2) R^j{}_i G^{00} + 2a_3 R^{0j}{}_i{}^0
\nonumber
\eea
and (\ref{main}) lead to the following conditions:
\be
2a_1+2a_2=0,  \qquad 2a_4+2a_5-1 =0 , \qquad
\det D_i{}^j \neq 0
\label{eqineq}
\ee
The last inequality may be violated in strong gravitational field and
as we comment below this fact may lead to causal problems.

Suppose that all the conditions (\ref{eqineq}) are fulfilled.
For simplicity we also choose $a_3=0$. Then we have the classical
action of the form (\ref{genact}) with the coefficients
\be
a_1=\frac{\xi_1}{2}, \quad
a_2=-\frac{\xi_1}{2}, \quad
a_3=0, \quad
a_4=\frac{1}{2}-\xi_2, \quad
a_5=\xi_2
\ee
where $\xi_1$, $\xi_2$ are two arbitrary coupling parameters.

One of the secondary constraints
\be
\varphi^{(2)}_0 = \n^\alpha E_{\alpha 0}
\ee
does not contain velocities $\dot H_{00}$, $\dot H_{0i}$ and so
its conservation leads to a new constraint
$\varphi^{(3)} \approx \n_0\n^\alpha E_{\alpha 0}$. After exclusion
from this expression the accelerations $\ddot H_{ij}$ we get this
constraint as the following combination of the equations of motion:
\bea
\varphi^{(3)} &=& \n_0\n^\mu E_{\mu0} - \xi_2 G_{00} R^{ij} E_{ij}
\nonumber\\&&{}
+\frac{1}{D-2} \Bigl[ m^2 G_{00} + (\xi_2-\xi_1)RG_{00} + R_{00}\Bigr]
G^{ij} E_{ij}
\eea
$\varphi^{(3)}$  contains neither the
acceleration $\ddot H_{00}$ nor the velocity $\dot H_{00}$. It means
that its conservation in time leads to another new constraints
\be
\varphi^{(4)}\approx \n_0 \varphi^{(3)}
\ee
and hence the total number of constraints is the
same as in the flat spacetime.

Unfortunately, analysis of causal properties of such
a theory on static background\cite{causality} shows that there can be
spacetime regions where some of the above constraints fail to be of
the second class and some components of $H_{\mu\nu}$ may propagate
with superluminal velocities.

Another possible way to fulfill the consistency requirements
(\ref{main}) is to consider spacetimes representing solutions
of vacuum Einstein equations with arbitrary cosmological
constant:
\be
R_{\mu\nu}=\frac{1}{D}G_{\mu\nu}R \; {.}
\label{einst}
\ee
In this case the coefficients $a_4$, $a_5$ in the lagrangian
(\ref{genact}) are absent and the elements of the matrix $\hat\Phi$
take the form:

\bea
A &=& R G^{00}(2a_1+2a_2-\frac{\d 1}{\d D})
\nonumber\\
B^j &=& R G^{0j}(2a_1+4a_2) + 2a_3R^{0j}{}_0{}^0 + m^2 G^{0j}
\nonumber\\
C_i &=&  0
\nonumber\\
D_i{}^j &=&    2a_3 R^{0j}{}_i{}^0 +
 G^{00} \delta_i^j (2a_1-\frac{\d 2}{\d D}) - m^2 G^{00} \delta_i^j
\eea
The simplest way to make the rank of such a matrix to be equal to
$D-1$ is provided by the following choice of the coefficients:
\be
2a_1 + 2a_2 -\frac{1}{D} = 0, \qquad a_3 = 0, \qquad
2R \biggl(a_1-\frac{1}{D}\biggr) - m^2 \neq 0 {.}
\ee

As a result, we have one-parameter family of theories:
\bea
&&
a_1 =\frac{\xi}{D}, \quad a_2 = \frac{1-2\xi}{2D},\quad
a_3=0,\quad a_4=0, \quad a_5 = 0
\nonumber\\&&
R_{\mu\nu}=\frac{1}{D}G_{\mu\nu}R , \qquad
\frac{2(1-\xi)}{D} R + m^2 \neq 0 {.}
\eea
with $\xi$ an arbitrary real number.

The action in this case takes the form
\bea
S&=&\int d^D x\sqrt{-G} \biggl\{ \frac{1}{4} \nabla_\mu H \nabla^\mu H
-\frac{1}{4} \nabla_\mu H_{\nu\rho} \nabla^\mu H^{\nu\rho}
-\frac{1}{2} \nabla^\mu H_{\mu\nu} \nabla^\nu H
\nonumber
\\&&{}\qquad
+\frac{1}{2} \nabla_\mu H_{\nu\rho} \nabla^\rho H^{\nu\mu}
+\frac{\xi}{2D} R H_{\mu\nu} H^{\mu\nu} +\frac{1-2\xi}{4D} R H^2
\nonumber
\\&&{}\qquad
- \frac{m^2}{4} H_{\mu\nu} H^{\mu\nu} + \frac{m^2}{4} H^2
 \biggr\} {.}
\label{curvact}
\eea
and the corresponding equations of motion are
\bea
E_{\mu\nu}&=&\nabla^2 H_{\mu\nu} - G_{\mu\nu} \nabla^2 H +
\nabla_\mu \nabla_\nu H
+ G_{\mu\nu} \nabla^\alpha \nabla^\beta H_{\alpha\beta}
- \nabla_\sigma \nabla_\mu H^\sigma{}_\nu
\nonumber
\\&&{}
- \nabla_\sigma \nabla_\nu H^\sigma{}_\mu
+ \frac{2\xi}{D} R H_{\mu\nu} + \frac{1-2\xi}{D} RH G_{\mu\nu}
\nonumber
\\&&{}
-m^2 H_{\mu\nu} + m^2 H G_{\mu\nu} = 0
\label{cons_eq}
\eea
The secondary constraints built out of them are
\be
\varphi^{(2)}_\mu=\nabla^\alpha E_{\alpha\mu} =
(\nabla_\mu H -\nabla^\alpha H_{\mu\alpha})
\biggl( m^2+\frac{2(1-\xi)}{D}R\biggr)
\label{seccon}
\ee
Just like in the flat case, in this theory the conditions
$\dot \varphi^{(2)}_i\approx 0$ define the accelerations $\ddot H_{0i}$
and the condition $\dot\varphi^{(2)}_0\approx 0$ after excluding
$\ddot H_{0i}$ gives a new constraint, i.e. the acceleration $\ddot
H_{00}$ is not defined at this stage.

To define the new constraint in a covariant form we use the
following linear combination of $\dot\varphi^{(2)}_\mu$, equations of
motion, primary and secondary constraints:
\bea
\varphi^{(3)} &=&
\frac{m^2}{D-2} G^{\mu\nu} E_{\mu\nu}
+ \nabla^\mu \nabla^\nu E_{\mu\nu}
+ \frac{2(1-\xi)}{D(D-2)} R G^{\mu\nu} E_{\mu\nu} =
\\{}
\nonumber
&=& H \frac{1}{D-2} \biggl( \frac{2(1-\xi)}{D} R + m^2 \biggr)
\biggl( \frac{D+2\xi(1-D)}{D} R + m^2 (D-1)\biggr)
\eea
This gives tracelessness condition for the field $H_{\mu\nu}$
provided that parameters of the theory fulfill the conditions:
\be
\frac{2(1-\xi)}{D} R + m^2 \neq 0 {,} \qquad
\frac{D+2\xi(1-D)}{D} R + m^2 (D-1) \neq 0
\label{ineq}
\ee

Requirement of conservation of $\varphi^{(3)}$ leads to one more
constraint
\bea
\dot \varphi^{(3)} \sim \dot H  \quad\Longrightarrow\quad
\varphi^{(4)} = \dot H \approx 0 {.}
\eea
The last acceleration $\ddot H_{00}$ is expressed from the
condition $\dot\varphi^{(4)}\approx 0$.

Using the constraints for simplifying the equations of motion we see
that the original equations are equivalent to the following system:
\bea
&&\nabla^2 H_{\mu\nu}
+ 2 R^\alpha{}_\mu{}^\beta{}_\nu H_{\alpha\beta}
+ \frac{2(\xi-1)}{D} R H_{\mu\nu} - m^2 H_{\mu\nu} = 0 {,}
\nonumber\\&&
H^\mu{}_\mu=0 {,} \qquad\qquad\dot H^\mu{}_\mu=0 {,} \qquad\qquad
\nabla^\mu H_{\mu\nu} = 0 {,}
\label{curv_shell}
\\&&
G^{00}\n_0\n_i H^i{}_\nu - G^{0i}\n_0\n_i H^0{}_\nu
- G^{0i}\n_i\n_0 H^0{}_\nu - G^{ij}\n_i\n_j H^0_\nu
\nonumber
\\&&{}\qquad\qquad
-2R^{\alpha0\beta}{}_\nu H_{\alpha\beta}
- \frac{2(\xi-1)}{D} RH^0{}_\nu + m^2 H^0{}_\nu = 0 {.}
\nonumber
\eea
The last expression represents $D$ primary constraints.

For any values of $\xi$ (except two degenerate values excluded by
(\ref{ineq})) the theory describes the same number of degrees of
freedom as in the flat case - the symmetric, covariantly transverse
and traceless tensor.  $D$ primary constraints guarantees
conservation of the transversality conditions in time.

Let us now consider the causal properties of the theory. Again, if we
tried to use the equations of motion in the original lagrangian form
(\ref{cons_eq}) then the characteristic matrix
would be degenerate. After having used the constraints we obtain the
equations of motion written in the form (\ref{curv_shell}) and the
characteristic matrix becomes non-degenerate:
\be
M_{\mu\nu}{}^{\lambda\kappa} (n) =
\delta_{\mu\nu}{}^{\lambda\kappa} n^2, \qquad
n^2 = G^{\alpha\beta} n_\alpha n_\beta {.}
\ee
The characteristic cones remains the same as in the flat case. At any
point $x_0$ we can choose locally
$G^{\alpha\beta}(x_0)=\eta^{\alpha\beta}$
and then
\be
\left. n^2 \right|_{x_0} = - n_0^2 + n_i^2
\ee
Just like in the flat case the equations are hyperbolic and causal.

Now let us discuss the massless limit of the theory under
consideration. There are several points of view on the
definition of masslessness in a curved spacetime of an arbitrary
dimension. We guess that the most physically accepted definition is
the one referring to appearance of a gauge invariance for some
specific values of the theory parameters
(see e.g.\cite{massless,nolland} for a recent discussion).

In our case it means that the real mass parameter $M$
for the field $H_{\mu\nu}$ in an Einstein spacetime is defined as
\be
M^2 = m^2 + \frac{2(1-\xi)}{D} R
\label{realm}
\ee
When $M^2=0$ instead of $D$ secondary constraints $\varphi_\mu^{(2)}$
we have $D$ identities for the equations of motion
$\n^\mu E_{\mu\nu}\equiv 0$ and the theory acquires gauge invariance
$\delta H_{\mu\nu}=\n_\mu \xi_\nu + \n_\nu \xi_\mu$. This explains
the meaning of the first condition in (\ref{ineq}), it just tells us
that the theory is massive.

In fact, two parameters $m^2$ and $\xi$
enter the action (\ref{curvact}) in a single combination $M^2$
(\ref{realm}). Since scalar curvature is constant in Einstein
spacetime there is no way to distinguish between the corresponding
terms $\sim \xi R HH$, $\sim m^2 HH$ (with arbitrary $\xi$, $m$) in
the action.  The difference between the two will appear only if we
consider Weyl rescaling of the metric. Note that the ``massless''
theory with $M^2=0$ is not Weyl invariant. In the case of dS/AdS
spacetimes the difference between masslessness, conformal and gauge
invariance and null cone propagation was discussed in detail
in\cite{desnep}.  In our case the theory obviously cannot possess Weyl
invariance.

The second inequality (\ref{ineq}) is more mysterious. If it fails to
hold, i.e. if
$
M^2 = M^2_c \equiv \frac{D-2}{D(D-1)} R
$
then instead of the constraint $\varphi^{(3)}$ the scalar identity
\be
\n^\mu\n^\nu E_{\mu\nu} + \frac{R}{D(D-1)} G^{\mu\nu} E_{\mu\nu} =0
\ee
with the corresponding gauge invariance
\be
\delta H_{\mu\nu} = \n_\mu \n_\nu \epsilon + \frac{R}{D(D-1)}
G_{\mu\nu} \epsilon
\label{strange}
\ee
arise.

Appearance of this gauge invariance with a scalar parameter was first
found for the massive spin 2 in spacetime of constant curvature
in\cite{desnep} and was further investigated\cite{higuchi,bengt} in
spacetimes with positive cosmological constant. Our analysis shows
that this gauge invariance is a feature of more general spin 2
theories in arbitrary Einstein spacetimes.  In this case we can
simplify the equations of motion using the secondary constraints
(\ref{seccon}):
\be
\n^2 H_{\mu\nu}- \n_\mu\n_\nu H
+ 2R_\mu{}^\alpha{}_\nu{}^\beta H_{\alpha\beta}
+ \frac{2-D}{D(D-1)} RH_{\mu\nu} -\frac{1}{D(D-1)}
RG_{\mu\nu} H = 0 {.}
\ee
After imposing the gauge condition\footnote{It does not fix
(\ref{strange}) completely and the residual symmetry
with the prameter obeying
$\bigl(\n^2+\frac{R}{D-1}\bigr)\epsilon=0$ remains.} $H=0$ one can see
that these equations describe causal propagation of the field
$H_{\mu\nu}$ but the number of propagating degrees of freedom
corresponds to neither massive nor massless spin 2 free field.
It was argued in\cite{higuchi,bengt} that appearance of the gauge
invariance (\ref{strange}) leads to such
pathological properties as  violation of the classical Hamiltonian
positiveness and negative norm states in the quantum version of the
theory. One should expect similar problems in the general
spin 2 theory in arbitrary Einstein spacetime described in this
paper.

\section{Consistent equations in arbitrary gravitational background}

In the previous section we analyzed a possibility of consistent
description of the spin 2 field on specific spacetime manifolds. Now
we will describe another possibility which allows to remove any
restrictions on the external gravitational background by means of
considering a lagrangian in the form of infinite series in inverse
mass $m$. Existence  of dimensionful mass parameter $m$ in the theory
let us construct a lagrangian with terms of arbitrary orders
in curvature multiplied by the corresponding powers of $1/m^2$, i.e
having the following schematic form:
\begin{eqnarray}&&
S_H =\int d^D x\sqrt{-G} \biggl\{ \nabla H \nabla H
+ R H H + m^2 H H
\nonumber\\&&\qquad{}
+\frac{1}{m^2} ( R \nabla H \nabla H + R H \nabla\nabla H + R R H H)
+ O\Bigl(\frac{1}{m^4}\Bigr)
 \biggr\}
\label{higher}
\end{eqnarray}
Actions of this kind are expected to arise naturally in string
theory where the role of mass parameter is played by string
tension $m^2=1/\alpha'$ and perturbation theory in $\alpha'$ will
give for background fields effective actions of the form
(\ref{higher}). Possibility of constructing consistent equations for
massive higher spin fields as series in curvature was recently
studied in\cite{klish2} where such equations were derived in
particular case of symmetrical Einstein spaces in linear in curvature
order.

Here we demonstrate that requirement of consistency with the
flat spacetime limit can be fulfilled perturbatively in $1/m^2$
for arbitrary gravitational background at least in the lowest order.
We use the same general scheme of calculating lagrangian
constraints as in the previous section. The only difference is that
each condition will be considered perturbatively and can be solved
separately in each order in $1/m^2$.

Primary constraints in the theory described by the action
(\ref{higher}) should be given by the equations $E^0{}_\mu\approx 0$.
Requirement of absence of second time derivatives in these equations
will give some restrictions on coefficients in higher orders in
$1/m^2$, for example, in terms like $R\n H\n H$.

Consistency with the flat spacetime limit requires existence of one
additional constraint among conservation conditions of the
secondary constraints. The
advantage of having a theory in the form of infinite series consists
in the possibility to calculate the determinant of the matrix
$\hat\Phi$
perturbatively in $1/m^2$. Assuming that the lower
right subdeterminant  of the matrix is not zero (it is not zero in
the flat case) one has
\be
\det \hat\Phi = (A-BD^{-1}C ) \det D  \; , \qquad
\det D \neq 0
\ee
Converting the matrix D perturbatively
\be
D^{-1} =  - \frac{1}{m^2 G^{00}} \delta_i^j +
O\left(\frac{1}{m^4}\right)
\ee
we get
\be
A-BD^{-1}C = R G^{00} 2(a_1+a_2) + R^{00} (2a_4+2a_5-1) +
O\left(\frac{1}{m^2}\right)
\ee
So consistency with the flat limit imposes at this order in $m^2$ two
conditions on the five non-minimal couplings in the lagrangian
(\ref{higher}) and we are left with a three parameters family of
theories:
\be
a_1=\frac{\xi_1}{2} , \quad
a_2=-\frac{\xi_1}{2} ,  \quad
a_3=\frac{\xi_3}{2}  ,  \quad
a_4=\frac{1}{2}-\xi_2 , \quad
a_5=\xi_2 .
\ee
The action (\ref{higher}) then takes the form:
\bea
&&
S_H =\int d^D x\sqrt{-G} \biggl\{ \frac{1}{4} \nabla_\mu H \nabla^\mu H
-\frac{1}{4} \nabla_\mu H_{\nu\rho} \nabla^\mu H^{\nu\rho}
-\frac{1}{2} \nabla^\mu H_{\mu\nu} \nabla^\nu H
\nonumber
\\&&
\qquad
{}+\frac{1}{2} \nabla_\mu H_{\nu\rho} \nabla^\rho H^{\nu\mu}
{}+\frac{\xi_1}{4} R H_{\alpha\beta} H^{\alpha\beta}
-\frac{\xi_1}{4} R H^2
+\frac{1-2\xi_2}{4} R^{\alpha\beta} H_{\alpha\sigma} H_\beta{}^\sigma
\nonumber\\&&\qquad{}
+\frac{\xi_2}{2} R^{\alpha\beta} H_{\alpha\beta} H
+\frac{\xi_3}{2} R^{\mu\alpha\nu\beta} H_{\mu\nu} H_{\alpha\beta}
- \frac{m^2}{4} H_{\mu\nu} H^{\mu\nu} + \frac{m^2}{4} H^2
\nonumber\\&&\qquad{}
+ O\Bigl(\frac{1}{m^2}\Bigr)
 \biggr\}
\label{consistent}
\eea

In this case the rank of the matrix $\hat\Phi$ is equal to $D-1$ and
one can construct from the conservation conditions for the secondary
constraints
\be
\n_0 \varphi^{(2)}_\nu = \n_0\n^\mu E_{\mu\nu} = \hat\Phi_\nu{}^\mu
\ddot H_{0\mu} + \ldots
\ee
one covariant linear combination which does not contain acceleration
$\ddot H_{0\mu}$:
\be
\varphi^{(3)} \approx \n^\mu \n^\nu E_{\mu\nu}
- \frac{1}{m^2} R^{\alpha\nu} \n_\alpha\n^\mu E_{\mu\nu}
+ O\Bigl(\frac{1}{m^4}\Bigr)
\label{phi3}
\ee

Derivatives of the field $H_{\mu\nu}$ enter this expression in such a
way that it does not contain the accelerations $\ddot H_{00}$,
$\ddot H_{0\mu}$  and the velocity $\dot H_{00}$. It means that just
like in the flat case the conservation condition
$\dot\varphi^{(3)}\approx 0$ leads to another new constraints
$\varphi^{(4)}$ and the last acceleration $\ddot H_{00}$ is defined
from $\dot\varphi^{(4)}\approx 0$. The total number of constraints
coincides with that in the flat spacetime.

The constraints $\varphi^{(3)}$ and $\varphi^{(2)}_\mu$ can be solved
perturbatively in $1/m^2$ with respect to the trace and the
longitudinal part of $H_{\mu\nu}$
\be
\varphi^{(3)} \sim H + O\Bigl(\frac{1}{m^2}\Bigr) , \qquad
\varphi^{(2)}_\mu \sim \n^\mu H_{\mu\nu} + O\Bigl(\frac{1}{m^2}\Bigr)
\label{nice}
\ee
and used for reducing the original equations of motion to the
conditions:
\bea
&& \n^2 H_{\mu\nu} - m^2 H_{\mu\nu} + \xi_1 RH_{\mu\nu}
- (\frac{1}{2}+\xi_2) (R_\mu{}^\alpha H_{\alpha\nu}
+ R_\nu{}^\alpha H_{\alpha\mu})
\nonumber\\&&{}
+(\xi_3+2) R_\mu{}^\alpha{}_\nu{}^\beta H_{\alpha\beta}
+ \frac{\xi_2-\xi_3-1}{D-1} G_{\mu\nu} R^{\alpha\beta} H_{\alpha\beta}
+ O\Bigl(\frac{1}{m^2}\Bigr) = 0 {,}
\nonumber\\&&
H + O\Bigl(\frac{1}{m^2}\Bigr) = 0 {,} \qquad
\n^\mu H_{\mu\nu} + O\Bigl(\frac{1}{m^2}\Bigr) = 0
\label{finhigh}
\eea
and also to the $D$ primary constraints $E^0{}_\mu$.
We see that even in this lowest order in $m^2$ not all
non-minimal terms in the equations are arbitrary. Consistency with
the flat limit leaves only three arbitrary parameters while the number
of different non-minimal terms in the equations is four.

However, if gravitational field is also subject to some dynamical
equations of the form $R_{\mu\nu}=O(1/m^2)$ then the system
(\ref{finhigh}) contains only one non-minimal coupling in the lowest
order
\bea
&& \n^2 H_{\mu\nu} - m^2 H_{\mu\nu}
+(\xi_3+2) R_\mu{}^\alpha{}_\nu{}^\beta H_{\alpha\beta}
+ O\Bigl(\frac{1}{m^2}\Bigr) = 0 {,}
\nonumber\\&&
H + O\Bigl(\frac{1}{m^2}\Bigr) = 0 {,} \qquad
\n^\mu H_{\mu\nu} + O\Bigl(\frac{1}{m^2}\Bigr) = 0 {,}
\nonumber\\&&
R_{\mu\nu} + O\Bigl(\frac{1}{m^2}\Bigr) = 0
\label{ricci}
\eea
and is consistent for any its value.

Requirement of causality does not impose any restrictions on the
couplings in this order. The characteristic matrix of (\ref{finhigh})
is non-degenerate, second derivatives enter in the same way as in the
flat spacetime, and hence the light cones of the field $H_{\mu\nu}$
described by (\ref{finhigh}) are the same as in the flat case.
Propagation is causal for any values of $\xi_1$, $\xi_2$, $\xi_3$. In
higher orders in $1/m^2$ situation becomes more complicated and we
expect that requirement of causality may give additional restrictions
on the non-minimal couplings.

Concluding this section we would like to stress once more that
the theory (\ref{consistent}) admits any gravitational background and
so no inconsistencies arise if one treats gravity as dynamical field
satisfying Einstein equations with the energy - momentum tensor for
the field $H_{\mu\nu}$. The action for the system of interacting
gravitational field and massive spin 2 field and the Einstein
equations for it are:
\begin{eqnarray}
&&S = S_E + S_H {,} \qquad
S_E = - \frac{1}{\kappa^{D-2}} \int d^D x \sqrt{-G} R {,}
\nonumber\\&&
R_{\mu\nu} - \frac{1}{2} G_{\mu\nu} R = \kappa^{D-2} T^H_{\mu\nu} {,}
\qquad
T^H_{\mu\nu} = \frac{1}{\sqrt{-G}}\frac{\delta S_H}{\delta G^{\mu\nu}}
\end{eqnarray}
with $S_H$ given by (\ref{consistent}). However, making the metric
dynamical we change the structure of the second derivatives by
means of nonminimal terms $\sim RHH$ which  can spoil causal
propagation of both metric and massive spin 2 field\cite{aragone}.
This will impose extra restrictions on the parameters of the
theory. Also, one can consider additional requirements the
theory should fulfill, e.g. tree level unitarity of graviton -
massive spin 2 field interaction\cite{cpd}.

\section{String theory in background of massive spin 2 field}

In this section we will consider sigma-model description of an open
string interacting with two background fields -- massless graviton
$G_{\mu\nu}$ and second rank symmetric tensor field $H_{\mu\nu}$ from
the first massive level of the open string spectrum. We will
show that effective equations of motion for these fields are of the
form (\ref{ricci}) and explicitly calculate the coefficient $\xi_3$ in
these equations in the lowest order in $\alpha'$.

Classical action has the form
\bea
S&=&S_0+S_I=
\nonumber\\&=&
\frac{1}{4\pi\alpha'}\int_M \!\!d^2z\sqrt{g}
      g^{ab}\partial_ax^\mu\partial_bx^\nu G_{\mu\nu}
+\frac{1}{2\pi\alpha'\mu}\int_{\partial M} e dt \;
      H_{\mu\nu}\dot{x}^\mu\dot{x}^\nu
\label{actstring}
\eea
Here $\mu,\nu=0,\ldots,D-1$; $a,b=0,1$ and we introduced the notation
$\dot{x}^\mu=\frac{dx^\mu}{edt}$. The first term $S_0$ is an integral
over two-dimensional string world sheet $M$ with metric $g_{ab}$ and
the second $S_I$ represents a one-dimensional integral over its
boundary with einbein $e$. We work in euclidian signature and
restrict ourselves to flat world sheets with straight boundaries. It
means that both two-dimensional scalar curvature and extrinsic
curvature of the world sheet boundary vanish and we can always choose
such coordinates that $g_{ab}=\delta_{ab}$, $e=1$.

Theory has two dimensionful parameters. $\alpha'$ is the fundamental
string length squared, $D$-dimensional coordinates $x^\mu$ have
dimension $\sqrt{\alpha'}$. Another parameter $\mu$ carries dimension
of inverse length in two-dimensional field theory (\ref{actstring})
and plays the role of renormalization scale. It is introduced in
(\ref{actstring}) to make the background field $H_{\mu\nu}$
dimensionless. In fact, power of $\mu$ is responsible for the number
of massive level to which a background field belongs because one
expects that open string interacts with a field from $n$-th massive
level through the term
$$
\mu^{-n} (\alpha')^{-\frac{n+1}{2}}
\int_{\partial M} e dt\; \dot{x}{}^{\mu_1} \ldots
\dot{x}{}^{\mu_{n+1}} H_{\mu_1 \ldots \mu_{n+1}} (x)
$$

The action (\ref{actstring}) is non-renormalizable from the point of
view of two-dimensional quantum field theory. Inclusion of
interaction with any massive background produces in each loop an
infinite number of divergencies and so requires an infinite number of
different massive fields in the action. But massive modes from the
$n-$th massive level give vertices proportional to $\mu^{-n}$ and so
they cannot contribute to renormalization of fields from lower
levels. Of course, this argument assumes that we treat the theory
perturbatively defining propagator for $X^\mu$ only by the term with
graviton in (\ref{actstring}). Now we will use such a scheme to
carry out renormalization of (\ref{actstring}) dropping all the terms
$O(\mu^{-2})$.

Varying (\ref{actstring}) one gets classical equations of motion with
boundary conditions:
\begin{eqnarray}
&& g^{ab} D_a \partial_b x^\alpha \equiv
g^{ab}(\partial_a \partial_b x^\alpha + \Gamma^\alpha_{\mu\nu}(G)
\partial_a x^\mu \partial_b x^\nu) = 0,
\nonumber\\&&
\left. G_{\mu\nu} \partial_n x^\mu \right|_{\partial M} -
\frac{2}{\mu} {\cal D}^2_t x^\mu H_{\mu\nu}
\nonumber\\&&{}\qquad\qquad
+\frac{1}{\mu}
{\dot x}^\mu {\dot x}^\lambda (\nabla_\nu H_{\mu\lambda}
-\nabla_\mu H_{\nu\lambda} - \nabla_\lambda H_{\mu\nu}) = 0
\label{class}
\end{eqnarray}
where  $\partial_n=n^a \partial_a$, $n^a$ -- unit inward normal
vector to the world sheet boundary and ${\cal D}^2_t x^\mu = {\ddot
x}^\mu+\Gamma^\mu_{\nu\lambda}(G) {\dot x}^\nu {\dot x}^\lambda$.

Divergent part of the one loop effective action has the form
\begin{eqnarray}
&&\Gamma^{(1)}_{div}=
    -\frac{\mu^{-\varepsilon-1}}{2\pi\varepsilon}\int_{\partial M}
    dt e(t) \dot{x}^\mu\dot{x}^\nu
   \left( \nabla^2 H_{\mu\nu} - 2R_\mu{}^\alpha H_{\alpha\nu}
+ R_\mu{}^\alpha{}_\nu{}^\beta H_{\alpha\beta} \right)
\nonumber\\&&\qquad{}
+\frac{\mu^{-\varepsilon}}{4\pi\varepsilon}
    \int_{M} d^{2+\varepsilon} z\sqrt{g}
    g^{ab}\partial_ax^\mu\partial_bx^\nu R_{\mu\nu}
+ O(\mu^{-2})
\end{eqnarray}
where the terms $O(\mu^{-2})$ give contributions to
renormalization of only the second and higher massive levels.
Hence one-loop renormalization of the background fields looks like:
\begin{eqnarray}
\stackrel{\circ}{G}_{\mu\nu}&=&\mu^\varepsilon
  G_{\mu\nu}-\frac{\alpha'\mu^\varepsilon}{\varepsilon}R_{\mu\nu}
\nonumber\\
\stackrel{\circ}{H}_{\mu\nu}&=&\mu^\varepsilon H_{\mu\nu}
  +\frac{\alpha'\mu^\varepsilon}{\varepsilon}
  \left( \nabla^2 H_{\mu\nu} -2 R^\sigma{}_{(\mu} H_{\nu)\sigma}
  + R_\mu{}^\alpha{}_\nu{}^\beta H_{\alpha\beta} \right)
\label{1loop}
\end{eqnarray}
with circles denoting bare values of the fields. We would like to
stress once more that higher massive levels do not influence the
renormalization of any given field from the lower massive levels and
so the result (\ref{1loop}) represents the full answer for
perturbative one-loop renormalization of $G_{\mu\nu}$ and
$H_{\mu\nu}$.

Now to impose the condition of Weyl invariance of the theory at the
quantum level we calculate the trace of energy momentum tensor
in $d=2+\varepsilon$ dimension:
\begin{equation}
  T(z) = g_{ab}(z) \frac{\delta S}{\delta g_{ab}(z)} =
\frac{\varepsilon\mu^{-\varepsilon}}{8\pi\alpha'}
g^{ab}(z)\partial_ax^\mu\partial_bx^\nu G_{\mu\nu}
-\frac{\mu^{-1-\varepsilon}}{4\pi\alpha'} H_{\mu\nu} \dot{x}^\mu
\dot{x}^\nu \delta_{\partial M}(z)
\end{equation}
and perform one-loop renormalization of the composite operators:
\begin{equation}
\bigl( \dot{x}^\mu \dot{x}^\nu \stackrel{\circ}{H}_{\mu\nu} \bigr)_0 =
\mu^{-\varepsilon}
\bigl[ \dot x^\mu \dot x^\nu  H_{\mu\nu} \bigr]
\end{equation}
\begin{eqnarray}&&
(g^{ab}\partial_a x^\mu \partial_b x^\nu
   \stackrel{\circ}{G}_{\mu\nu})_0 =\mu^\varepsilon
   \Bigl[g^{ab}\partial_a x^\mu \partial_b x^\nu (G_{\mu\nu}
   -\frac{\alpha'}{\varepsilon} R_{\mu\nu})\Bigr]
\\&&{}
+\frac{\alpha'\mu^{-1+\varepsilon}}{\varepsilon}
    \left[H_{\alpha}{}^\alpha\delta_{\partial M}''(z)
   + {\cal D}^2_tx^\mu (\nabla_\mu H_\alpha{}^\alpha
   - 4 \nabla^\alpha H_{\alpha\mu}) \delta_{\partial M}(z)
\right.
\nonumber\\&&{}
\left.
   + {\dot x}^\mu {\dot x}^\nu (
    \nabla_\mu \nabla_\nu H_\alpha{}^\alpha
   - 4 \nabla^\alpha \nabla_{(\mu} H_{\nu)\alpha}
   + 2 \nabla^2 H_{\mu\nu} - 2 R_\mu{}^\alpha{}_\nu{}^\beta
     H_{\alpha\beta} ) \delta_{\partial M}(z) \right]
\nonumber
\end{eqnarray}
Here delta-function of the boundary $\delta_{\partial M}(z)$ is
defined as
\begin{eqnarray}
&& \int_M\delta_{\partial M}(z)
V(z)\sqrt{g(z)} d^2z =\int_{\partial M}V|_{z\in\partial M} e(t) dt
\end{eqnarray}

The renormalized operator of the energy momentum tensor
trace is:
\begin{eqnarray}
8\pi[T]&=&{}
- \Bigl[ g^{ab} \partial_a x^\mu \partial_b x^\nu
E_{\mu\nu}^{(0)}(x)\Bigr] + \frac{2}{\mu} \delta_{\partial M}(z)
   \Bigl[ {\dot x}^\mu {\dot x}^\nu E^{(1)}_{\mu\nu}(x) \Bigr]
\nonumber\\&&{}
+ \frac{1}{\mu} \delta_{\partial M}(z)
   \Bigl[ {\cal D}^2_t x^\mu E^{(2)}_\mu(x) \Bigr]
+ \frac{1}{\mu} \delta''_{\partial M}(z) \Bigl[ E^{(3)}(x) \Bigr]
\end{eqnarray}
where
\begin{eqnarray}
E^{(0)}_{\mu\nu} (x)&=& R_{\mu\nu} + O(\alpha')
\nonumber\\
E^{(1)}_{\mu\nu}(x)&=&
\nabla^2 H_{\mu\nu} - \nabla^\alpha \nabla_\mu H_{\alpha\nu}
- \nabla^\alpha \nabla_\nu H_{\alpha\mu}
\nonumber\\&&{}
- R_\mu{}^\alpha{}_\nu{}^\beta H_{\alpha\beta}
+ \frac{1}{2} \nabla_\mu \nabla_\nu H_\alpha{}^\alpha
- \frac{1}{\alpha'} H_{\mu\nu}  + O(\alpha')
\nonumber\\
E^{(2)}_\mu(x)&=& \nabla_\mu H_\alpha{}^\alpha
- 4 \nabla^\alpha H_{\alpha\mu} + O(\alpha')
\nonumber\\
E^{(3)}(x)&=&H_\alpha{}^\alpha + O(\alpha')
\label{E}
\end{eqnarray}
Terms of order $O(\alpha')$ arise from the higher loops contributions.

The requirement of quantum Weyl invariance tells that all $E(x)$
in (\ref{E}) should vanish and so they are interpreted as effective
equations of motion for background fields. They contain vacuum
Einstein equation for graviton (in the lowest order in $\alpha'$),
curved spacetime generalization of the mass shell condition for the
field $H_{\mu\nu}$ with the mass $m^2=(\alpha')^{-1}$ and $D+1$
additional constraints on the values of this fields and its first
derivatives. Taking into account these constraints and the Einstein
equation we can write our final equations arising from the Weyl
invariance of string theory in the form:
\begin{eqnarray}
&& \nabla^2 H_{\mu\nu}
+ R_\mu{}^\alpha{}_\nu{}^\beta H_{\alpha\beta}
- \frac{1}{\alpha'} H_{\mu\nu} + O(\alpha') = 0 {,}
\nonumber\\&&
\nabla^\alpha H_{\alpha\nu} + O(\alpha') =0 {,} \qquad
H^\mu{}_\mu +O(\alpha') =0 {,}
\nonumber\\&&
R_{\mu\nu} +O(\alpha')= 0 {.}
\label{final}
\end{eqnarray}
They coincide with the equations found in the previous section
(\ref{ricci}) with the value of non-minimal coupling $\xi_3=-1$.

In fact, Einstein equations should not be vacuum ones
but contain dependence on the field $H_{\mu\nu}$ through its energy -
momentum tensor $T^H_{\mu\nu}$. Our calculations could not produce
this dependence because such dependence is expected to arise only if
one takes into account string world sheets with non-trivial topology
and renormalizes new divergencies arising from string loops
contribution\cite{fs}.
\be
R_{\mu\nu} +O(\alpha')= T^H_{\mu\nu} - \frac{1}{D-2}
 T^H{}^\alpha{}_\alpha {,}
\ee
where explicit form of the lowest contributions to the
energy-momentum tensor $T^H_{\mu\nu}$ can be determined
only from sigma model on world sheets with topology of annulus.

In order to determine whether the equations (\ref{final}) can be
deduced from an effective lagrangian (and to find this lagrangian)
one would need two-loop calculations in the string sigma-model.
Two-loop contributions to the Weyl anomaly coefficients $E^{(i)}$  are
necessary because the effective equations of motion
(\ref{E},\ref{final}) are not the equations directly following from a
lagrangian but some combinations of them similar to (\ref{finhigh}).
In order to reverse the procedure of passing from the original
lagrangian equations to (\ref{finhigh}) one would need the next to
leading contributions in the conditions for $\nabla^\mu H_{\mu\nu}$
and $H_{\mu\nu}$ (\ref{final}).

\section*{Acknowledgments}

We are grateful to our collaborators D.~Gitman and V.~Krykhtin, and
also to S.~Kuzenko, H.~Osborn, B.~Ovrut, A.~Tseytlin, M.~Vasiliev
and G.~Veneziano for useful discussions. The work was supported by
GRACENAS grant, project 97-6.2-34; RFBR grant, project 99-02-16617;
RFBR-DFG grant, project 99-02-04022 and INTAS grant N 991-590. I.L.B.
is grateful to FAPESP for support of the research.

\end{document}